\documentclass{cjour}
 \usepackage{epsfig}

\begin{document}

 \renewcommand\dbltopfraction{1.0}
 \renewcommand\textfraction{0.1}
 \renewcommand\floatpagefraction{1.0}

 \authorrunninghead{A. van Heukelum, G. T. Barkema, R. H. Bisseling}
 \titlerunninghead{DNA electrophoresis studied with the cage model}

 \title{DNA electrophoresis studied with the cage model}

 \author{A. van Heukelum\thanks{A.vanHeukelum@phys.uu.nl} and 
 G. T. Barkema\thanks{G.T.Barkema@phys.uu.nl}}
 \affil{Institute for Theoretical Physics, Utrecht University,\\
 Princetonplein 5, 3584 CC Utrecht, The Netherlands}
 \and
 \author{R. H. Bisseling\thanks{Rob.Bisseling@math.uu.nl}}
 \affil{Mathematical Institute, Utrecht University,\\
 PO Box 80010, 3508 TA Utrecht, The Netherlands}


 \date{May 31, 2002}

 \abstract{The cage model for polymer reptation, proposed by Evans and
Edwards, and its recent extension to model DNA electrophoresis, are
studied by numerically exact computation of the drift velocities for
polymers with a length $L$ of up to 15 monomers. The computations show
the Nernst-Einstein regime ($v \sim E$) followed by a regime where the
velocity decreases exponentially with the applied electric field
strength. In agreement with de Gennes' reptation arguments, we find that
asymptotically for large polymers the diffusion coefficient $D$
decreases quadratically with polymer length; for the cage model, the
proportionality coefficient is $DL^2=0.175(2)$. Additionally we find
that the leading correction term for finite polymer lengths scales as
$N^{-1/2}$, where $N=L-1$ is the number of bonds.}

 \begin{article}

 \section{Introduction}

 In the rapidly-growing fields of molecular genetics and genetic
engineering, gel electrophoresis is a technique of great importance. One
reason is that it enables efficient separation of polymer strands by
length. In DNA electrophoresis, strands of DNA of various lengths are
injected into a gel composed of agarose and a buffer solution. Since DNA
is acidic, it becomes negatively charged. Next, an electric field is
applied which causes the DNA to migrate in one direction. Since shorter
strands travel faster than longer ones, the initial mixture of strands
will become separated, allowing the measurement of the relative
concentrations of strands of different lengths, or the isolation of
strands with a particular length. Given the great practical importance
of DNA electrophoresis, there is much interest in gaining an
understanding of precisely what the mechanisms of gel electrophoresis
are and how the migration rate depends on strand length, applied
electric field, and the properties of the agarose gel.

 It is known that in the gel, agarose forms long strands which
cross-link and impede movement of the polymer transverse to its length;
its movement is dominated by a mechanism which de
Gennes~\cite{deGennes71} has dubbed {\it reptation}: movement of a
polymer along its own length by diffusion of stored length.

 A commonly used lattice model to simulate the dynamics of reptation is
the so-called ``repton model'', introduced by Rubinstein in
1987~\cite{Rubinstein87}. Rubinstein had already conjectured that the
diffusion coefficient $D$ as a function of polymer length $L$ for long
polymers is given by $DL^2=1/3$ (with large finite-size effects); this
conjecture was further corroborated by Van Leeuwen and
Kooiman~\cite{LK92,KL93a,KL93b}, and finally proven by Pr\"ahofer and
Spohn~\cite{Prahofer96}. The repton model has been extended to study
electrophoresis by Duke~\cite{Duke89,Duke90a,Duke90b}, and the resulting
model---known as the Duke-Rubinstein model---has been studied
numerically and analytically by several
groups~\cite{WVD91,BMW94,DSV92,BN97a,BN97b} and compared to
experiments~\cite{BCM96}.

 The main findings of these studies are that the property $DL^2=1/3$
(with large finite-size effects) in combination with the
fluctuation-dissipation theorem results in a drift velocity $v\sim E/L$
for small electric field strength $E$, and that for some value of $E\sim
1/L$ this regime crosses over in a regime where the drift velocity
ceases to be length dependent and is given by $v \sim E^2$, the
so-called {\it plateau mobility} regime. Furthermore, it was found to be
a property of the model in the limit of large $E$, that the drift
velocity decays exponentially, $v \sim e^{(2-L)E/2}$ and $v \sim
e^{(3-L)E/2}$ for even and odd $L$, respectively~\cite{Kolomeisky98}.

 Before the introduction of the repton model, Evans and Edwards had
introduced the so-called ``cage model'' to simulate the dynamics of
reptation~\cite{Evans81}. Also in this model, $DL^2$ approaches a
constant in the limit of large chains~\cite{bk98}, which in combination
with the fluctuation-dissipation theorem leads to $v \sim E/L$ for small
electric field strengths. This model has recently been extended to
electrophoresis and studied with Monte Carlo
simulations~\cite{Heukelum2000}. Besides the expected
fluctuation-dissipation regime, these simulations also featured the
plateau mobility regime where $v \sim E^2$; these are the two regimes
that were identified for the Duke-Rubinstein model using Monte Carlo
simulations. Additionally, a third regime was reported where $v$
decreases with increasing $E$.

 This article presents numerically exact computations on the cage model,
extended for electrophoresis as in Ref.~\cite{Heukelum2000}. As in most
models, numerically exact results can only be obtained for relatively
small systems (here, for polymers up to a length of $L=15$), but they do
not have the inherently large statistical errors of Monte Carlo results.
Thus, they allow for a different class of analysis techniques, for
instance those exploiting numerical differentiation. The combination of
numerically exact results for short chains with the Monte Carlo results
for larger chains reported in Refs.~\cite{bk98,Heukelum2000} provides a
more complete picture of the model.

 The calculations done in this work, with a chain length of up to
$L=15$, are computationally challenging, and could only be obtained by
the exploitation of symmetries in the model, combined with the
application of parallel processing. The state vector of the cage model
for electrophoresis has $6^{L-1}$ components, and the original
transition matrix which represents the transition probabilities between
polymer configurations is of size $6^{L-1} \times 6^{L-1}$ (see
Section~\ref{sec:model}). We show that many components of the
steady-state vector are equal, because the configurations they belong to
are equivalent. By using those equivalences, the original transition
matrix could be reduced significantly (see Section~\ref{sec:exploit}).
The parallel implementation of the computation is done by spreading the
nonzeros of the sparse transition matrix over the processors.
Interprocessor communication is reduced by exploiting the specific
sparsity structure of the matrix (see Section~\ref{sec:parproc}). The
combined effect of decreasing the matrix size, improving the
eigenspectrum of the matrix, and applying parallel processing
accelerates the computation by more than a factor of a million, allowing
us to reach larger values of $L$; in this paper we present numerically
exact values for the diffusion coefficients for polymers up to length
$L=15$ (see Section~\ref{sec:physconc}). We also present computation
times and parallel efficiency results for up to 64 processors of a Cray
T3E computer (see Section~\ref{sec:compconc}). The conclusions are
summarized in Section~\ref{sec:conc}.

 \section{Cage model for reptation}
 \label{sec:model}

\begin{figure}
 \begin{center}
 \includegraphics[width=3.25in]{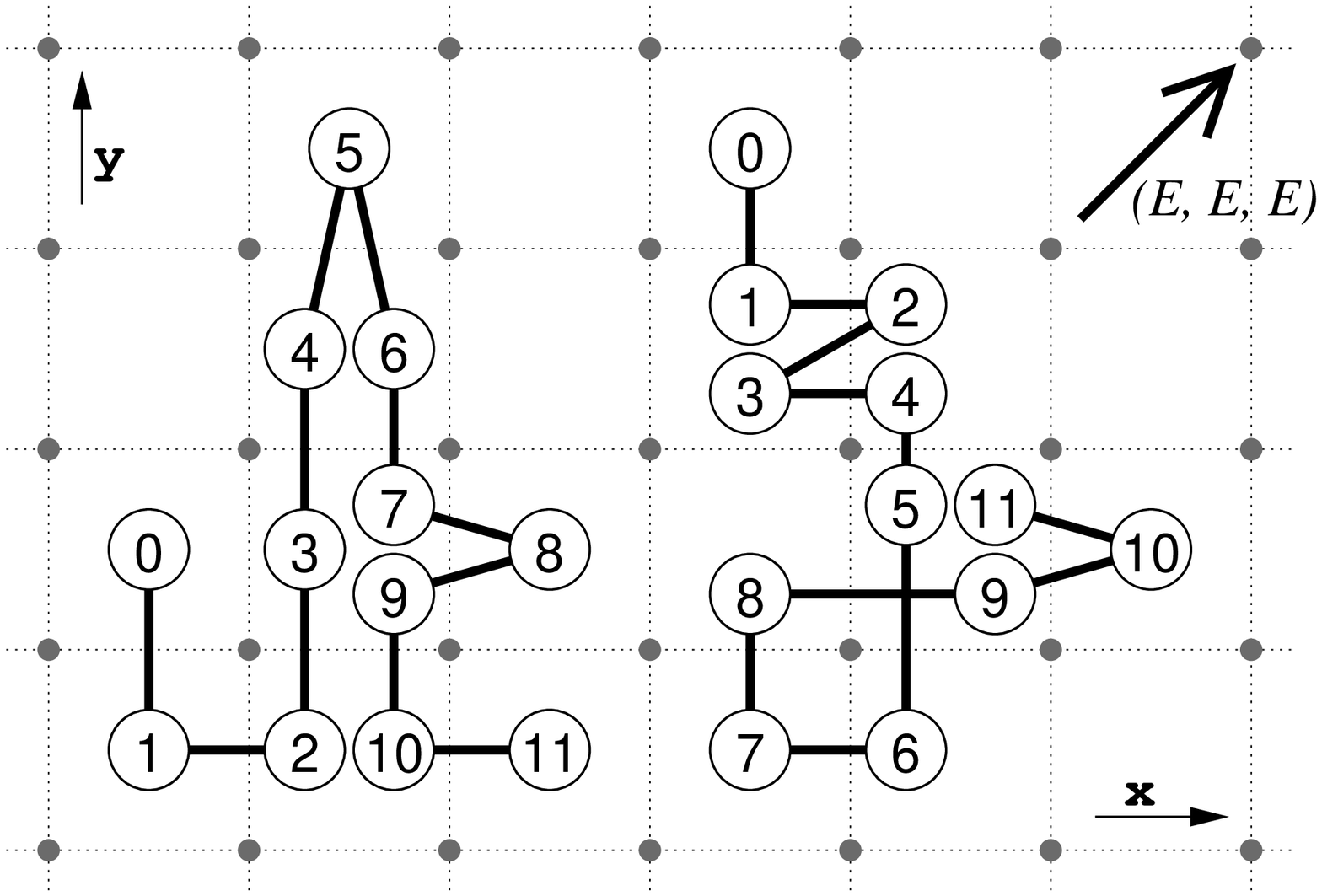}
 \end{center}
 \caption{The cage model. The dotted lines denote the gel strands in the
$x$-$y$ plane, and the large gray dots are the gel strands in the $z$
direction. The space between the gel strands represents the pores of the
gel. The polymer is modeled as a chain of monomers; two adjacent
monomers reside in nearest neighbor pores. We denote bonds that are
going right, left, up, down, out of the paper, and into the paper by
{\tt +x}, {\tt -x}, {\tt +y}, {\tt -y}, {\tt +z}, and {\tt -z}
respectively. The two example configurations were chosen to be planar,
for clarity. The electric field vector points diagonally out of the
paper.}
 \label{cageexample}
 \end{figure}

 The cage model, introduced by Evans and Edwards~\cite{Evans81},
describes a polymer that moves through a gel. The polymer is modeled as
a chain of ``monomers'', connected by {\em bonds}. Two monomers
connected by a bond must reside in adjacent sites of a cubic lattice. No
other excluded volume interactions are enforced, so each lattice site
may contain many monomers. Figure~\ref{cage3dex} shows an impression of
the model. The configuration of a cage polymer is most easily defined by
the set of directions of all the bonds. The bond representations shown
in Fig.~\ref{cageexample} are {\tt
-y\,+x\,+y\,+y\,+y\,-y\,-y\,+x\,-x\,-y\,+x} and {\tt
-y\,+x\,-x\,+x\,-y\,-y\,-x\,+y\,+x\,+x\,-x}. A part of a configuration
consisting of a monomer with two opposite bonds is called a {\em kink}.
In the left part of Fig.~\ref{cageexample}, the configuration features
kinks at monomers 5 and 8, and in the right part at 2, 3, and 10.
 \begin{figure}
 \begin{center}
 \includegraphics[width=3.25in]{cage3d.eps}
 \end{center}
 \caption{An impression of the cage model for reptation. The polymer
consists of a sequence of monomers, connected by unit-length bonds.}
 \label{cage3dex}
 \end{figure}

 The gel is modeled by the edges of a cubic lattice, translated by a
vector $(\frac12, \frac12, \frac12)$ relative to the lattice on which
the polymer resides. The dynamics of a cage polymer consists of those
single monomer moves for which the polymer does not cut gel strands.
This leaves two classes of allowed moves: (i) a kink is randomly
replaced by a kink in one of the six possible directions; (ii) a bond at
an end monomer is randomly replaced by a bond in one of the six possible
directions. Every other single monomer move is forbidden because it
would cause the polymer to cross a gel strand. The model does not impose
self-avoidance on the polymer chain, which is justified because a
typical mixture of long DNA strands is semidilute in electrophoresis
experiments, and because for short strands the self-avoidance is
negligible since the persistence length of DNA is much larger than its
diameter. The diffusion coefficient of the polymer chains can be
computed from the displacement of the center of mass~\cite{bk98}.

 The cage model has been extended to include the effects of an electric
field on the motion of (charged) polymers~\cite{Heukelum2000}. The
possible transitions are the same, but the rates are different. The
electric field is $\vec{E}=(E, E, E)$, such that replacing a kink with
one of the three forward-pointing kinks (along the electric field)
occurs with rate $e^{qE}$, and replacing with one of the three
backward-pointing kinks (against the electric field) occurs with rate
$e^{-qE}$, where $q$ is the dimensionless charge of the moved monomer.
In the remainder of this paper, we assume that $q=1$.

 The set of all probabilities of the $6^{L-1}$ possible configurations
can be represented by a $6^{L-1}$-dimensional vector. The dynamics of
the model is then specified by a sparse $6^{L-1} \times 6^{L-1}$ matrix
$T$. The transition matrix $T$ has $\bigl[5(\frac{L-2}{6}+2)+1\bigr]
6^{L-1}$ nonzero elements: each polymer has $L-2$ inner monomers that
can move if their bonds are in opposite directions, and two end monomers
that can always move; a monomer that can move goes to one of five new
positions or the polymer stays unaltered.

 If the polymer is moved along the applied field from configuration $j$
to configuration $i$, then $T_{ij} = \delta t \cdot e^{E}$. If the
polymer is moved against the applied field, $T_{ij} = \delta t \cdot
e^{-E}$. The variable $\delta t$ can be interpreted as the time step in
Monte Carlo simulations. The diagonal element $T_{ii}$ is such that the
sum of each column is exactly one. An upper bound for the sum of the
off-diagonal elements in column $j$ is $\delta t\cdot 3L(e^E+e^{-E})$,
because at most $L$ kinks can move, each in at most three forward and
three backward directions. We choose $\delta t=(3L(e^E+e^{-E}))^{-1}$,
so that all elements in column $j$ are in the range $[0, 1]$. Thus,
$T_{ij}$ is the probability to move from configuration $j$ to $i$.  
Because we have a nonzero probability to stay in the same state,
$T_{ii}>0$ holds for all $i$. This implies $T_{ij}<1$ for all $i\neq j$.  
Because end monomers can always move, we also have $T_{ii}<1$.

 The cage model with an applied electric field is ergodic, i.e., every
configuration can reach every other configuration in a finite number of
steps. The steady-state vector $\vec a$ is the eigenvector of $T$ with
eigenvalue one (which is the largest eigenvalue), normalized such that
$\sum_i a_i = 1$. The drift velocity of the polymer along one of the
principal axes is
 \begin{equation}\label{eq1}
 v = \frac23 \sum_i a_i \bigl(b_i e^{E} - f_i e^{-E}\bigr)\mbox{,}
 \end{equation}
 where $b_i$ is the number of kinks and end monomers of polymer
configuration $i$ pointing backward (which can move forward with a rate
of $e^{E}$), and $f_i$ the number of kinks and end monomers pointing
forward. The factor of $2/3$ appears because moves occur along each of
the three principal axes, and because each kink move increases or
decreases the sum of the coordinates of a configuration by two.

 \section{Exploiting symmetries of the model}
 \label{sec:exploit}

 In the model that we study here, the electric field is chosen in the
$(1,1,1)$ direction, and consequently polymer configurations that are
related through rotation around the direction $(1,1,1)$ are equivalent,
i.e., their probability is the same, irrespective of the field strength.
Moreover, in many cases it is possible to rotate {\it part} of the
polymer around this direction while preserving this equivalence. If
polymer configurations are grouped into classes containing only
equivalent polymers, it is sufficient to determine the probability for
one polymer configuration per class rather than for all polymer
configurations, since by definition the probabilities are equal within a
class.

 Instead of working in the state space of all polymer configurations, we
work in the state space of all equivalence classes. Since equivalence
classes can easily contain thousands of configurations, the state space
is thus reduced by several orders of magnitude, and a tremendous speedup
is obtained. Next, we discuss how to identify whether two polymer
configurations are equivalent, and how physical quantities such as the
velocity can be computed within this reduced state space of equivalence
classes.

\begin{figure}
 \begin{center}
 \includegraphics[width=3.25in]{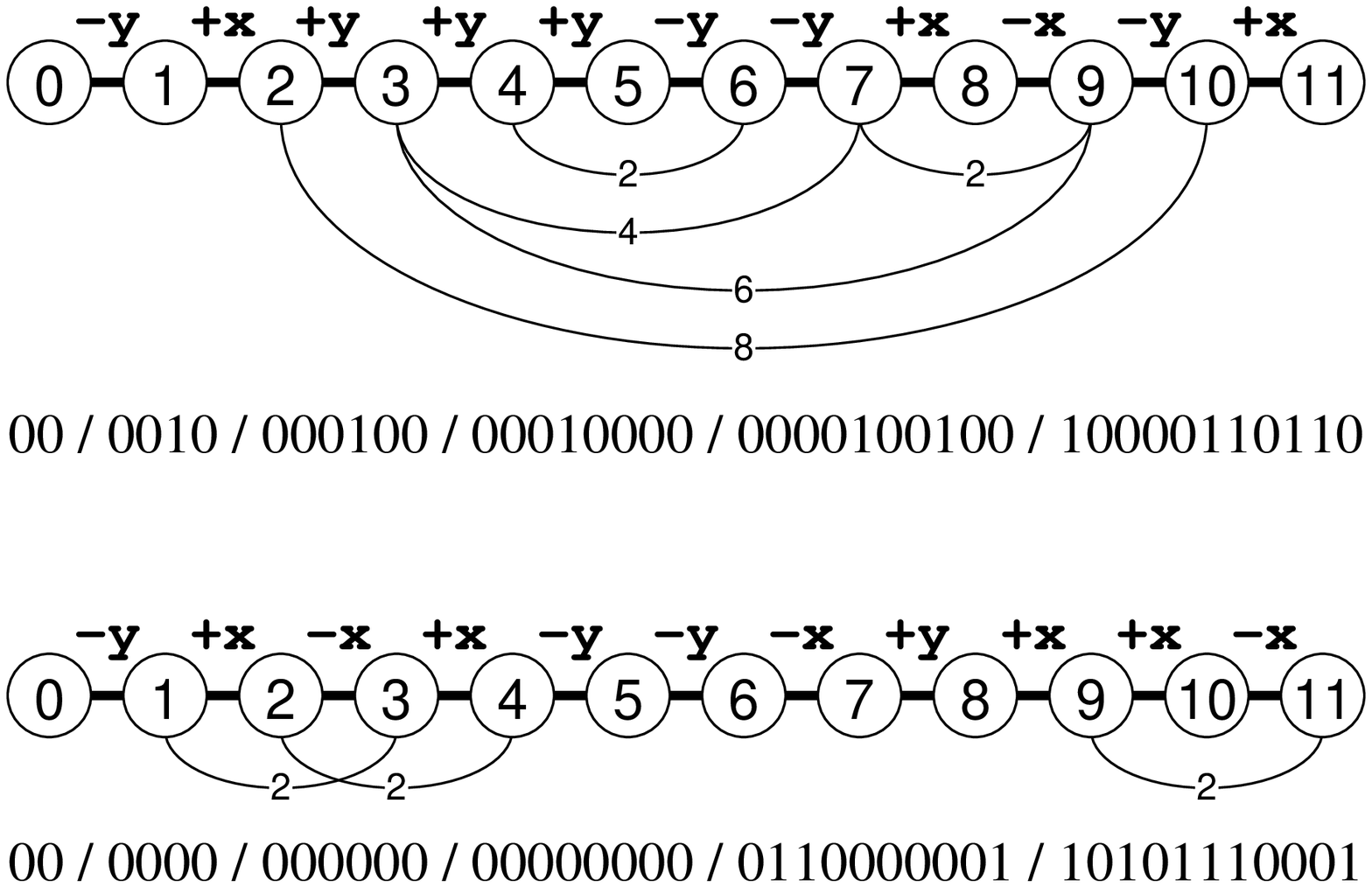}
 \end{center}
 \caption{Kink representations for the two examples from
Fig.~\ref{cageexample}. The arcs show which parts can be removed by
repeatedly removing kinks. The kink representations are also given as a
binary value; the slashes separate the removable parts of length $10$,
$8$, $6$, $4$, and $2$, and the bond directions. A bit 1 at position $r$
for part length $l$ means that the part between monomers $r$ and $r+l$
can be removed.}
 \label{kinkexample}
 \end{figure}

 To identify which polymer configurations are equivalent, we construct a
representation that puts equivalent configurations in the same class. We
call part of a configuration between two monomers {\em removable} if all
monomers between them can be removed by repeatedly removing kinks. A
kink is removed by deleting the central monomer and the two bonds
connected to it, and merging the two monomers adjacent to the central
monomer. In the left part of Fig.~\ref{cageexample}, monomers 4 and 6
are merged when the kink at monomer 5 is removed. If two polymers have
the same sequence of forward and backward bonds, and the same set of
removable pairs of monomers, they have the same {\it kink
representation}. The construction of such a representation is
illustrated in Fig.~\ref{kinkexample}. The kink representation gives a
unique number to each symmetry class. We can prove that all polymer
configurations with the same kink representation have the same
probability in the steady state (see Appendix) and we have verified this
explicitly up to $L\leq 9$. Furthermore, the forward/backward symmetry
was removed by also computing the kink representation starting at the
other end of the polymer, and then using only the one with the lower
binary value.

 The reduced state space is constructed by computing the kink
representation for each polymer configuration and removing the
duplicates (in our implementation, by using hashing). During this phase
some additional information is stored about each kink representation:
each bond representation that introduces a new kink representation is
stored along with the kink representation, and the total number of bond
representations for each kink representation is recorded. 
Table~\ref{tabsizes} shows the reduction of the configuration space
obtained by removing the symmetries. The kink representations are
enumerated by sorting them based on their binary value, with the
rightmost bit the least significant. This ordering has the property that
in most cases moves cause only small changes in binary values, e.g.,
replacing a kink {\tt +x\,-x} with {\tt -y\,+y} swaps two bond-direction
bits; replacing {\tt +x\,-x} with {\tt +y\,-y} even keeps them the same;
the removable-parts bits can be affected as well, but this becomes less
likely with increasing part length.
 \begin{table}
 \begin{center}
 \begin{tabular}{rrrrr}
 \hline
    &kink repre-&reduction&nonzero &reduction\\
 $L$&sentations &factor &elements&factor\\
 \hline
  3 &           5 &       7 &           19 &      22\\
  4 &           9 &      24 &           49 &      56\\
  5 &          37 &      35 &          233 &      75\\
  6 &          93 &      84 &          785 &     142\\
  7 &         340 &     137 &       3\,084 &     229\\
  8 &      1\,015 &     276 &      11\,003 &     407\\
  9 &      3\,534 &     475 &      41\,594 &     680\\
 10 &     11\,397 &     884 &     150\,645 &  1\,182\\
 11 &     39\,082 &  1\,547 &     559\,722 &  1\,999\\
 12 &    130\,228 &  2\,786 &  2\,032\,536 &  3\,451\\
 13 &    445\,315 &  4\,888 &  7\,479\,343 &  5\,869\\
 14 & 1\,505\,785 &  8\,674 & 27\,130\,349 & 10\,110\\
 15 & 5\,154\,859 & 15\,202 & 99\,199\,551 & 17\,248\\
 \hline
 \end{tabular}
 \end{center}
 \caption{The number of kink representations for polymer lengths
$L=3$--$15$, the reduction factor of the state space, the number of
nonzero elements for the matrix in the kink representation, and the
reduction factor of the number of nonzero elements.}
 \label{tabsizes}
 \end{table}

 The reduced transition matrix $T'$ is constructed one column at a time. 
For column $j$, we consider the possible moves of the bond
representation stored with kink representation $j$. For each resulting
bond representation, we compute the associated kink representation $i$,
concluding that kink representation $j$ can move to kink representation
$i$ in a single monomer move, and the reduced transition matrix element
$T'_{ij}$ is incremented by either $\delta t\cdot e^{E}$ or $\delta
t\cdot e^{-E}$, depending on the direction of the move.
Table~\ref{tabsizes} shows the resulting reduction factor in the number
of nonzero matrix elements. By construction, the matrix $T'$ has the
following properties. The sum of the elements in a column is one. All
elements are in the range $[0, 1)$; elements $T_{ii}$ are even in $(0,
1)$. The reduced model is ergodic, and its steady-state vector is the
normalized eigenvector with eigenvalue one (which is the largest). The
drift velocity is again computed by Eq.~(\ref{eq1}), where $a_i$ is now
the probability of class $i$, and $b_i$ ($f_i$) the number of backward
(forward)-pointing kinks and end monomers of a single polymer
configuration in class $i$.

 Repeatedly multiplying a starting vector by the reduced transition
matrix yields the steady-state vector. This iterative method to find the
eigenvector of the largest eigenvalue is well-known as the power method,
which converges if one eigenvalue is larger in absolute value than all
the others. The convergence of the power method depends on the ratio
between the largest and the second largest eigenvalue. We can compute
the eigenvector faster by applying the power method to $T^{(\omega)} =
I+\omega(T'-I)$, which has eigenvalues $\lambda^{(\omega)}_i = 1 +
\omega (\lambda'_i - 1)$, where the $\lambda'_i$ are the eigenvalues of
$T'$, and which has the same eigenvectors. We used $\omega = 2$, which
works well in practice.

 \section{Parallel processing approach}
 \label{sec:parproc}

 The reduced transition matrix for $L=15$ contains about $10^8$
elements, so both computational cost (10 Tflops for $50\,000$
iterations) and memory requirements (1.6 Gbyte) are too high for regular
workstations or PCs. We used the parallel programming library
BSPlib~\cite{Hill98} to obtain our results on a Cray T3E supercomputer,
using up to 64 processors. Within the Bulk Synchronous Parallel ({\em
BSP}) computing model~\cite{Valiant90}, computations and interprocessor
communications are separated by global synchronizations. BSPlib supports
two types of communication: direct remote memory access ({\em DRMA}) and
bulk synchronous message passing ({\em BSMP}). The DRMA operation {\bf
put} copies data into the memory space of a remote process at the next
synchronization, and {\bf get} retrieves data from a remote process at
the next synchronization. The BSMP operation {\bf send} sends a packet
to a queue on a remote processor, which, after the next synchronization,
can be accessed there with the {\bf move} operation. In total, the BSP
library has 20 primitives. We use the most efficient primitive, {\bf
put}. This can be done because the matrix remains constant during all
the iterations, so that it becomes worthwhile to analyze the
communication pattern beforehand and store a list of memory addresses to
be used as the target of {\bf put} operations.

 In our problem, for $L>12$, we cannot afford to store the complete
matrix on a single processor, so we need to distribute it over a number
of processors. The traditional way to do this is to distribute blocks of
rows of the matrix over the processors (even though for dense matrices
and certain sparse matrices it has been shown that this is not the most
efficient way for communication~\cite{Bisseling94}). In principle, we
use a more general, two-dimensional matrix distribution, which we will
tailor to our problem. The general computation of a matrix-vector
product ${\vec x}'=A \vec x$ with communication is as follows. The
matrix and vector are distributed over the processors: the nonzero
matrix elements $A_{ij}$ and the vector components $x_i$ are each
assigned to a processor. The matrix-vector product is given by $x'_i =
\sum_j A_{ij} x_j$. The first step is to communicate the components
$x_j$ to the processors with the corresponding $A_{ij}$. Now, each
processor $q$ computes the partial row sums $s_{i q}=\sum'_j A_{ij}
x_j$, where $\sum'_j$ denotes a summation that runs only over indices
$j$ for which $A_{ij}$ has been assigned to processor $q$. The partial
row sums are then communicated to the processor containing $x'_i$, and
finally they are accumulated into the components $x'_i$.

 The matrix we have to deal with is sparse and we exploit this in our
computations, since we only handle nonzero elements $A_{ij}$. In
addition, the nonzero structure shows ``patches'' with many nonzero
elements. We can exploit this to make our communications faster.
Consider a rectangular patch (i.e., a contiguous submatrix). A value
$x_j$ must be sent to the owner of the patch if an element $A_{ij}$ in
column $j$ of the patch is nonzero. It is likely that most columns of
the patch have at least one nonzero, so we might as well send all $x_j$
for that patch. This makes it possible to send a contiguous subvector of
$\vec x$, which is more efficient than sending separate components; this
comes at the expense of a few unnecessary communications. The trade-off
can be shifted by increasing or decreasing the patch size.

 To find suitable patches, we first divide the state vector into
contiguous subvectors. We use a heuristic to partition the matrix into
blocks of rows with approximately the same number of nonzeros. If we use
$P$ processors, and we want each processor to have $K$ subvectors, we
have to divide the vector into $KP$ subvectors. (The factor $K$ is the
overpartitioning factor.) This initial division tries to minimize the
computation time. Next, we adjust the divisions to reduce communication:
a suitable patch in the matrix corresponds to an input subvector of kink
representations where only the last few bits differ, and also to an
output subvector with that property. Therefore, we search for a pair of
adjacent kink representations that has a different bit as much as
possible to the left. This is a suitable place to split. We try to keep
the distance from the starting point as small as possible.

\begin{figure}
 \begin{center}
 \includegraphics[width=4.6in]{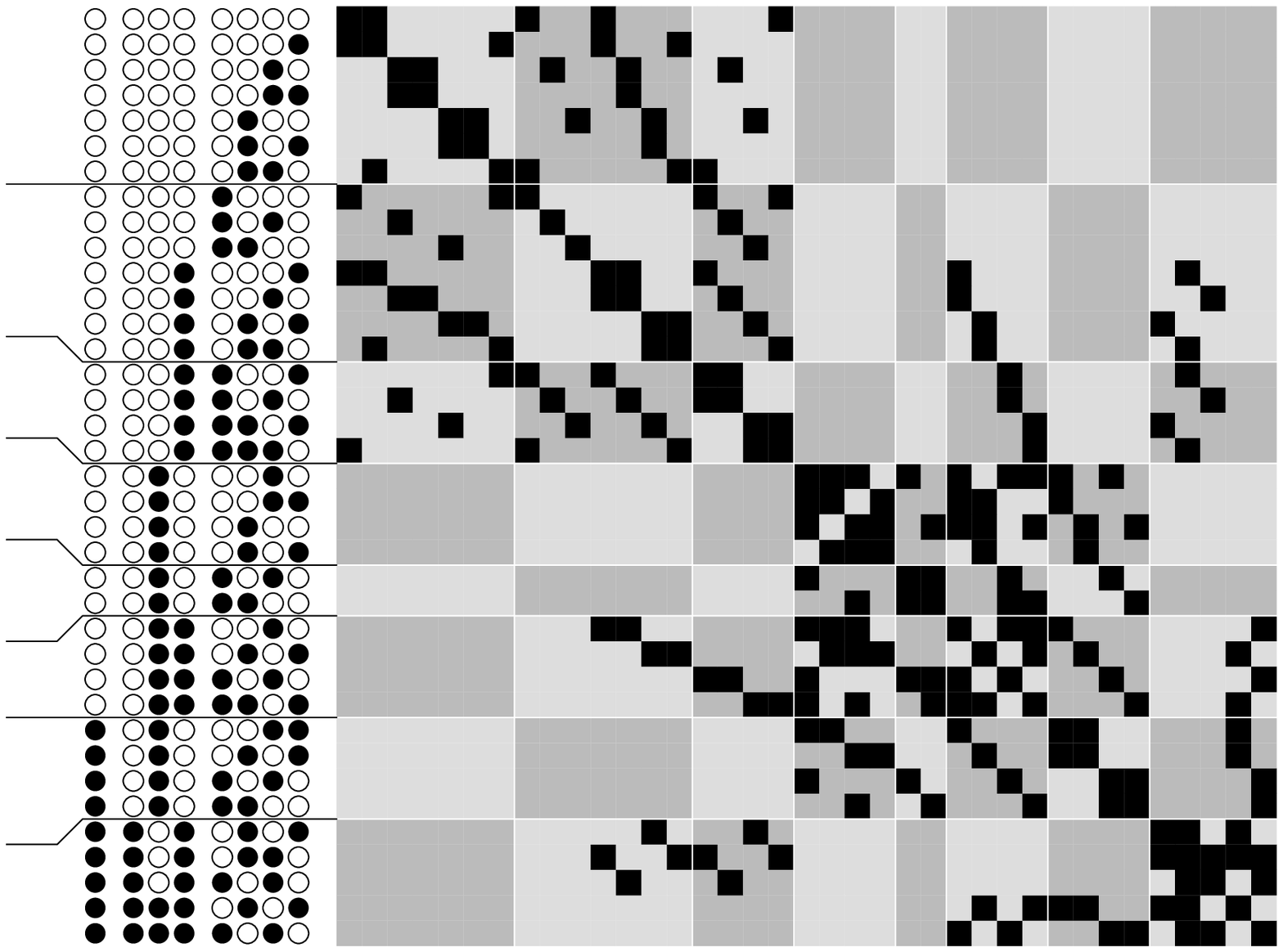}
 \end{center}
 \caption{Reduced transition matrix for polymer length $L=5$. The size
of the matrix is $37 \times 37$ and it has 233 nonzero elements, shown
as black squares. To the left of each row is the corresponding kink
representation written as a binary number, with black circles denoting 1
and open ones 0. The horizontal lines on the left show the initial
division of the reduced state vector into eight contiguous parts,
optimized to balance the number of nonzeros in the corresponding matrix
rows. The jumps of these lines indicate slight adjustments to make the
division fit the nonzero structure of the matrix. The resulting vector
division induces a division of the rows and columns of the matrix, and
hence a partitioning into 64 submatrices, shown by the gray checkerboard
pattern. Complete submatrices are now assigned to the processors of a
parallel computer.}
 \label{matexample}
 \end{figure}

\begin{figure}
 \begin{center}
 \includegraphics[width=4.9in]{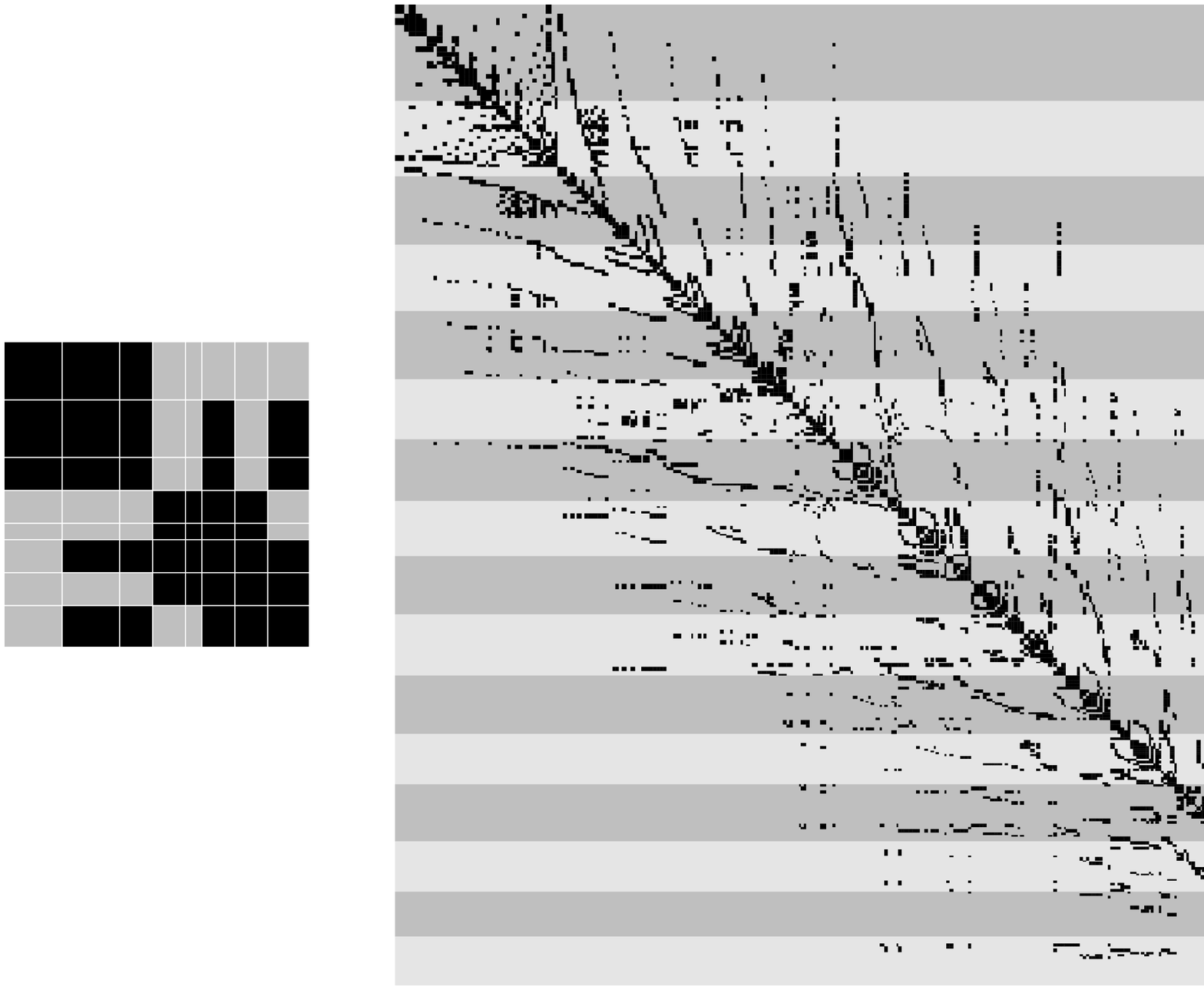}
 \end{center}
 \caption{Communication matrix for $L=5$ (left) and $L=13$ (right). Note
that the matrix for $L=5$ can be obtained by replacing each nonempty
submatrix in Fig.~\ref{matexample} by a single nonzero element. The
communication matrix for $L=13$, of size $320 \times 320$, is
distributed over $16$ processors in a row distribution.}
 \label{matcomm}
 \end{figure}

 As an example of the structure of the reduced transition matrices and
the division into submatrices, we show the nonzero structure of the
matrix for $L=5$ in Fig.~\ref{matexample} and its corresponding
communication matrix in Fig.~\ref{matcomm} (left). The communication
matrix is built from the partitioned transition matrix, by considering
each submatrix as a single element. It is a sparse matrix of much
smaller size which determines the communication requirements. Our
communication matrix for $L=13$ is given in Fig.~\ref{matcomm}
(right).

 \section{Results: drift velocities and diffusion~coefficients}
 \label{sec:physconc}

\begin{figure}
 \begin{center}
 \includegraphics[width=3.25in]{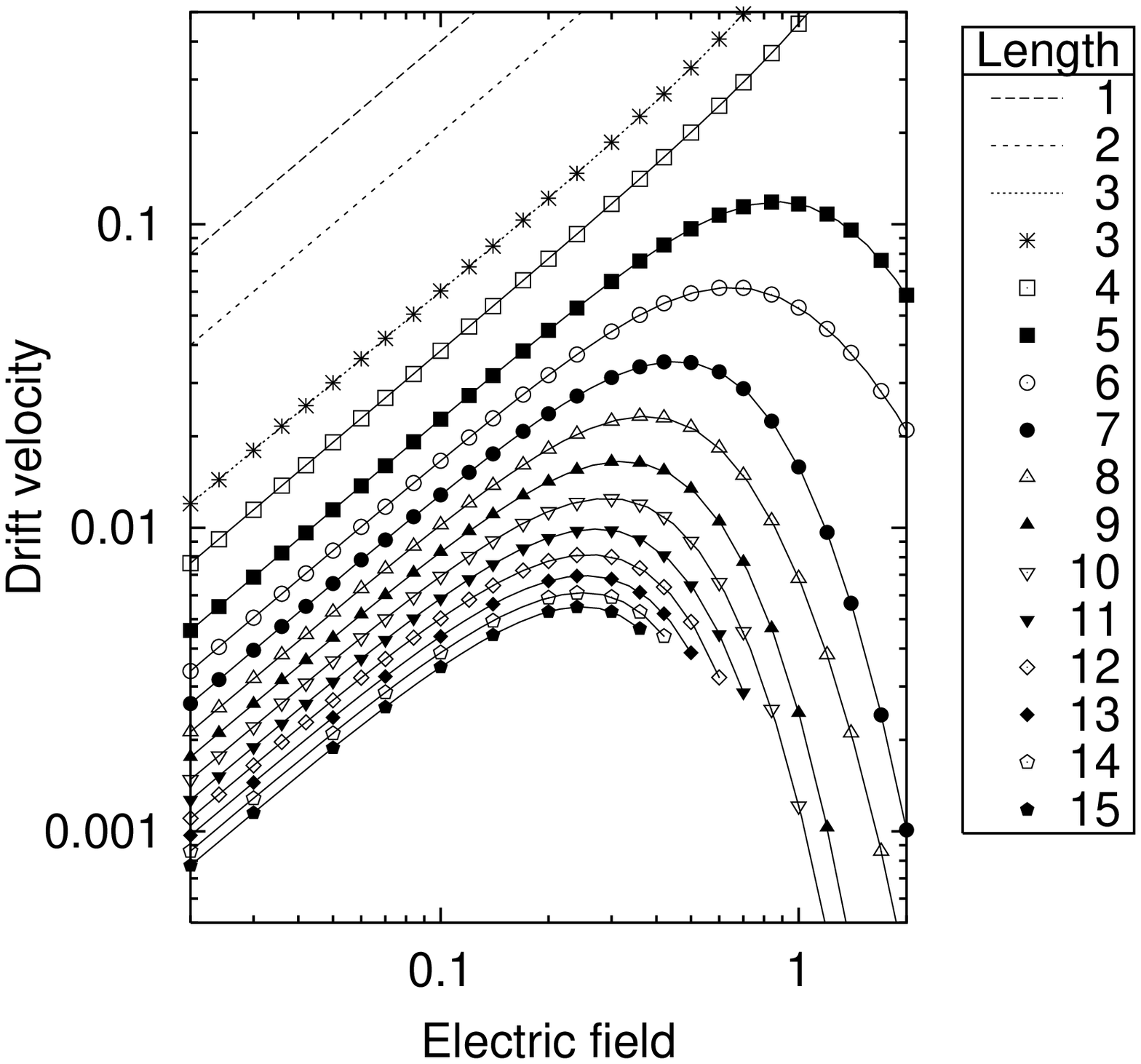}
 \end{center}
 \caption{The graphs show the computed drift velocities of the cage
polymers as a function of electric field strength $E$. For $E \leq 1$,
the relative error is less than $10^{-10}$; all other points have a
relative error less than $10^{-4}$. The graphs for lengths 1, 2, and 3
are $v_1=2(e^E-e^{-E})$, $v_2=e^E-e^{-E}$, and
$v_3=4(e^{3E}-e^{-3E})/(18+11(e^{2E}+e^{-2E}))$ respectively; for $L>3$,
the computed points are connected by straight lines.}
 \label{driftvel}
 \end{figure}

 Figure \ref{driftvel} shows the numerically exact values for the drift
velocity of the cage polymers up to length $L=15$. As expected,
initially the drift velocity increases linearly with field strength, and
eventually it reaches a maximum drift velocity, after which it decreases
exponentially with field strength. Clearly visible in Monte Carlo
data~\cite{Heukelum2000} is a regime just before the maximum velocity
where the drift velocity increases quadratically with the field
strength; in the numerically exact data presented here, for relatively
short chains, this regime is invisible.

 The diffusion coefficient is computed from the drift velocities using
the Nernst-Einstein relation $v=qLED$, which holds for vanishing $E$. We
used the computed velocities in the range $E=10^{-6}$ to $10^{-3}$ to
fit a second order polynomial to the mobility $\mu = v/E$ of the
polymers (see Table \ref{diffcoef}). The relative statistical error in
the mobility found by the fitting procedure was about $10^{-9}$.
 \begin{table}
 \begin{center}
 \begin{tabular}{rll}
 \hline
 $L$&$D$&$L^2D$\\
 \hline
  3 & 0.200\,000\,000\,000 & 1.800\,000\,000\,0\\
  4 & 0.095\,541\,401\,266 & 1.528\,662\,420\,3\\
  5 & 0.045\,892\,037\,845 & 1.147\,300\,946\,1\\
  6 & 0.028\,134\,332\,038 & 1.012\,835\,953\,4\\
  7 & 0.018\,844\,680\,457 & 0.923\,389\,342\,4\\
  8 & 0.013\,302\,014\,727 & 0.851\,328\,942\,5\\
  9 & 0.009\,776\,090\,804 & 0.791\,863\,355\,1\\
 10 & 0.007\,424\,928\,047 & 0.742\,492\,804\,7\\
 11 & 0.005\,790\,292\,327 & 0.700\,625\,371\,6\\
 12 & 0.004\,615\,107\,027 & 0.664\,575\,411\,8\\
 13 & 0.003\,746\,569\,186 & 0.633\,170\,192\,5\\
 14 & 0.003\,089\,624\,043 & 0.605\,566\,312\,4\\
 15 & 0.002\,582\,785\,984 & 0.581\,126\,846\,5\\
 \hline
 \end{tabular}
 \end{center} 
 \caption{Diffusion coefficients for cage polymers up to length $L=15$,
obtained by a second order polynomial fit to the mobility for field
strengths in the range $[10^{-6}, 10^{-3}]$. An upper bound for the
relative error is $10^{-9}$. For long polymers, $L^2D$ converges to a
constant~\protect\cite{bk98}.}
 \label{diffcoef}
 \end{table}

 It is known that asymptotically for large polymers the diffusion
coefficient behaves as $D \sim L^{-2}$, but with large finite-size
corrections for usual polymer lengths. As the polymers are modeled as a
random walk of $N=L-1$ steps, finite-size corrections of the order of
$N^{-1/2}$ are expected. Let us call $d(N)=D\cdot (N+1)^2=DL^2$, and
$d_\infty=(DL^2)_{L \to \infty}$; we expect that for large but finite
polymers $d(N) = d_\infty + a N^{-x}$. The parameters $a$ and $x$ can be
found from this equation by differentiation: $\frac{\partial d}{\partial
N} = -axN^{-x-1} \approx \frac12 \bigl(d(N+1)-d(N-1) \bigr)$. A
least-squares fit of the derivative of the new data against $N$ for
$N=8$--$13$ gives $a=2.469(5)$ and $x=0.512(6)$, strongly suggesting
finite-size corrections with an exponent $\frac12$. This shows an
advantage of the numerically exact computations over Monte Carlo
simulations in that we can compute the derivative of the data reliably.

 We used our new diffusion coefficients, combined with data from
Refs.~\cite{bk98,Heukelum2000} to find the length dependence of the
diffusion coefficient. A least-squares fit with
$d(N)=a+bN^{-1/2}+cN^{-1}$ gives $d(N)=0.172(6) + 0.63(8) N^{-1/2} +
3.3(2) N^{-1}$, and a least-squares fit with
$d(N)=(a'+b'N^{-1/2}+c'N^{-1})^{-1}$ gives $d(N)=(5.67(5)-22.2(5)
N^{-1/2} + 28(2) N^{-1})^{-1}$. Both of these expansions converge,
within the error margins, to the same value for large $N$. The first
expansion converges to 0.172(6), and the second expansion converges to
$1/5.67(5) = 0.176(2)$. Combining these results, we conclude that for
large $L$ the diffusion coefficient is $D=0.175(2) L^{-2}$. Our
diffusion coefficient agrees with that of Barkema and
Krenzlin~\cite{bk98}, but they reported a different finite-size scaling:
$DN^2=0.173+1.9N^{-2/3}$.

 \section{Results: Computation time and efficiency}
 \label{sec:compconc}

 Our computations were performed on a Cray T3E computer. The peak
performance of a single node of the Cray T3E is 600 Mflop/s for
computations. The bsp\_probe benchmark shows a performance of 47 Mflop/s
per node~\cite{Hill98}. The peak interprocessor bandwidth is 500 Mbyte/s
(bidirectional). The bsp\_probe benchmark shows a sustained
bidirectional performance of 94 Mbyte/s per processor when all 64
processors communicate at the same time. This is equivalent to a BSP
parameter $g=3.8$, where $g$ is the cost in flop time units of one
64-bit word leaving or entering a processor. The measured global
synchronization time for 64 processors is 48 $\mu$s, which is equivalent
to $l=2259$ flop time units.

 Table \ref{tab:speed} presents the execution time of one iteration of
the algorithm in two forms: the BSP cost $a+bg+cl$ counts the flops and
the communications and thus gives the time on an arbitrary computer with
BSP parameters $g$ and $l$, whereas the time in milliseconds gives the
measured time on this particular architecture, split into computation
and communication time. (The total measured synchronization time is
negligible.) The BSP cost can be used to predict the run time of our
algorithm on different architectures. Table \ref{tab:speed} also gives
the efficiency and speedup relative to a sequential program.
 \begin{table}
 \begin{center}
 \begin{tabular}{rrcccc}
 \hline
 $L$&$P$&BSP cost&time (ms)&efficiency&speedup\\
 \hline  
 12 &  8 &  545156 +   64716$g$ + 2$l$ &  47 + 4.3 & 85\% &  6.8\\
 13 & 16 & 1002824 +  187347$g$ + 2$l$ &  89 +  13 & 81\% & 13.0\\
 14 & 32 & 1836920 +  425152$g$ + 2$l$ & 169 +  44 & 73\% & 23.4\\
 15 & 64 & 3452776 + 1380415$g$ + 2$l$ & 330 + 112 & 67\% & 42.9\\
 \hline
 \end{tabular}
 \end{center}
 \caption{BSP cost, time, efficiency, and speedup for one matrix-vector
multiplication.}
 \label{tab:speed}
 \end{table}

 Peak computation performance is often only reached for dense
matrix-matrix multiplication; the performance for sparse matrix-vector
multiplication is always much lower. Comparing the flop count and the
measured computation time for the largest problem $L=15$, we see that we
achieve about 10.5 Mflop/s per processor. Comparing the communication
count with the measured communication time, we obtain a $g$-value of 8.1
$\mu$s, (or $g=3.8$ flop units; see above). This means that we attain
the maximum sustainable communication speed. This is due to the design
of our algorithm, which communicates contiguous subvectors instead of
single components. Furthermore, the results show that our choice to
optimize mainly the computation (by choosing a row distribution) is
justified for this architecture: the communication time is always less
than a third of the total time. For a different machine, with a higher
value of $g$, more emphasis must be placed on optimizing the
communication, leading to a two-dimensional distribution.

 Each iteration of our computation contains one matrix-vector
multiplication. The number of iterations needed for convergence depends
on the length of the polymer, and on the applied electric field. The
iteration was stopped when either the accuracy was better than
$10^{-10}$, or the number of iterations exceeded $100\,000$. In the
latter case, the accuracy was computed at termination. Typically, for
$L=15$ and a low electric field strength, $50\,000$ iterations are
needed. Only computed values with accuracy $10^{-4}$ or better are shown
in Fig.~\ref{driftvel}. For $L=12$, we compared the output for the
parallel program with that of the sequential program and found the
difference to be within rounding errors. The total speedup for $L=15$,
compared to a naive implementation (for which one would need 38.5~Tbyte
of memory), is a factor $1.5 \times 10^6$: a factor of $17\,248$ by
using a reduced state space, a factor of 2 by shifting the eigenvalues
of the reduced transition matrix, and a factor $42.9$ by using a
parallel program on $64$ processors.

 \section{Conclusions}
 \label{sec:conc}
 
 In numerically exact computations on the cage model, extended for the
study of DNA electrophoresis, we exploited symmetries of the model,
improved the eigenspectrum of the transition matrix, and applied
parallel processing. This has resulted in a computational speedup factor
of over a million.

 Regarding the cage model, we conclude that the polymer diffusion
coefficient $D$ scales asymptotically for large polymers as
$DL^2=0.175(2)$, in qualitative agreement with de Gennes' reptation
arguments, and in quantitative agreement with earlier simulation
reports~\cite{bk98}. The finite-size corrections are found to be a
combination of $N^{-1/2}$ (which asymptotically is the dominant
correction) and $N^{-1}$, and probably higher-order corrections; this is
in disagreement with earlier reports~\cite{BMW94,bk98} where the leading
corrections were reported to be $N^{-2/3}$, but in agreement with recent
density matrix renormalisation group computations by Carlon {\em et
al.}~\cite{CarlonUnp}.

 \appendix{Correctness proof of kink~representation
approach}\label{appendix}

 Our aim is to prove that all polymer configurations with the same kink
representation have the same probability in the steady state.  It is
sufficient to show that two configurations with the same kink
representation can move to the same set of six kink representations with
the moving of a certain kink or end monomer. We prove this by giving a
procedure for determining the resulting six kink representations.

 First, we introduce our notation. Define $R(i,j) $ as the statement
``the part of the configuration between monomers $i$ and $j$ is
removable'', where $0 \leq i, j < L$. (By this definition, $R(i,i)$
holds.) Define $S(i,j)$ as ``monomers $i$ and $j$ are at the same
site''. Define $\mathrm{sign}(i)=1$ if bond $[i,i+1]$ is in the
direction of the electric field, and $\mathrm{sign}(i)=-1$ otherwise.
We have the following useful properties.
 \begin{enumerate}
 \item $R(i,j)$ implies $S(i,j)$ and $j-i$ even.
 \item\label{prop:anykink} Let $i<j<k$. If $R(i,k)$ and $j$ is the
center of a kink, then the part between $i$ and $k$ can be removed
starting with the kink at $j$. Proof: By induction on the length of the
part.
 \item The relations $R$ and $S$ are equivalence relations between
monomers, i.e., they are reflexive, symmetric, and transitive. Proof:
Trivial, except for the proof of the transitivity of $R$, which uses the
previous property. For example, let $i<j<k$. If $R(i,k)$ and $R(i,j)$,
then a removal of the part $[j,k]$ can be obtained by starting the
removal of $R(i,k)$ by removing kinks in $[i,j]$.
 \item Let $j$ be the smallest integer such that $j>i$ and $R(i,j)$.
Then $R(i+1,j-1)$. Proof: By induction on the length of $[i,j]$.
 \item Let $i,i',j,j'$ be monomers with $|i-i'|=|j-j'|=1$. If $R(i,j)$
and $S(i',j')$, then $R(i',j')$. Proof: We treat the case $i'=i+1$ and
$j'=j+1$ as an example. First, we extend the part $[i,j']$ with a dummy
monomer $i-1$ at the site of $i'$. We can remove $[i-1,j']$ by first
removing $[i,j]$ and then removing the remaining kink $[i-1,j']$. By
Property~\ref{prop:anykink} above, we can also start with kink
$[i-1,i']$ and then remove $[i',j']$. Hence $R(i',j')$.
 \end{enumerate}

 Now assume that the kink at $i$ of a given polymer configuration moves.
(Moves of end monomers can be treated similarly.) We present a procedure
for generating the resulting six kink representations, which is based
solely on the original kink representation, i.e., on the relation $R$
and the bond signs. The correctness proof of this procedure uses the
properties above; for brevity, we omit the details. A kink exists at $i$
if and only if $R(i-1,i+1)$. In that case, $\mathrm{sign}(i) =
-\mathrm{sign}(i-1)$. The set of removable parts $[x,y]$ with $x,y \neq
i$ does not change; changes can only occur if $x=i$ or $y=i$. The
procedure checks for all $j$ whether $R(i+1,j)$. If so, monomer $i$ can
move to the sites of monomers $j-1$ and $j+1$, provided these monomers
exist. This is because $j-1$, $j+1$, and $i$ are all at distance one
from the site of $j$. If $j-1=i$, then $R(i+1,j)$ holds, and the move to
$j-1$ is the identity move, which does not change the kink
representation. Assume the move is to $j-1$ (the case $j+1$ is similar).
Assume $j-1 \neq i$. The new set of $x \neq i $ with $R(i,x)$ equals the
old set of $x \neq i$ with $R(j-1,x)$. The new $\mathrm{sign}(i)$ equals
the old $\mathrm{sign}(j-1)$.

 The generated moves are collected and duplicates are removed by using
the old relation $R$. For example, if $R(i+1,j)$ and $R(i+1,j')$ and we
have to check whether moves to $j-1 \neq i$ and $j'-1 \neq i$ are
identical, i.e., whether $S(j-1,j'-1)$, we can do this by checking the
old $R(j-1,j'-1)$. The total number of moves after duplicate removal is
at most six. To make the total six, extra moves are added. This is done
such that three moves have $\mathrm{sign}(i)=1$ and the others
$\mathrm{sign}(i)=-1$. The relation $R$ after such an extra move is the
same as before the moves, except that $R(i,x)$ becomes false for all $x
\neq i$. Note that $R(i,x)$ with $x>i$ implies that there exists a
smallest $x'>i$ with $R(i,x')$, and this in turn implies $R(i+1,x'-1)$,
so that the corresponding move of $i$ to $x'$ must have been generated
previously.

 \section*{Acknowledgments}

 We would like to thank the Dutch National Computer Facilities
foundation and the High Performance Applied Computing center at Delft
University of Technology for providing access to a Cray T3E.

 \end{article}


\begin{references}
 \frenchspacing
 \bibitem{deGennes71}
 P. G. de Gennes,
 Reptation of a polymer chain in the presence of fixed obstacles,
 {\it J. Chem. Phys.} {\bf55}, 572 (1971).
 \bibitem{Rubinstein87}
 M. Rubinstein,
 Discretized model of entangled-polymer dynamics,
 {\it Phys. Rev. Lett.} {\bf59}, 1946 (1987).
 \bibitem{LK92}
 J. M. J. van Leeuwen and A. Kooiman,
 The drift velocity in the Rubinstein-Duke model for electrophoresis,
 {\it Physica A} {\bf184}, 79 (1992).
 \bibitem{KL93a}
 A. Kooiman and J. M. J. van Leeuwen,
 Reptation models for electrophoresis,
 {\it Physica A} {\bf194}, 163 (1993).
 \bibitem{KL93b}
 A. Kooiman and J. M. J. van Leeuwen,
 The drift velocity in reptation models for electrophoresis,
 {\it J. Chem. Phys.} {\bf99}, 2247 (1993).
 \bibitem{Prahofer96}
 M. Pr\"ahofer and H. Spohn,
 Bounds on the diffusion constant for the Rubinstein-Duke model of
electrophoresis,
 {\it Physica A} {\bf233}, 191 (1996).
 \bibitem{Duke89}
 T. A. J. Duke,
 Tube model of field-inversion electrophoresis,
 {\it Phys. Rev. Lett.} {\bf62}, 2877 (1989).
 \bibitem{Duke90a}
 T. A. J. Duke,
 Monte-Carlo reptation model of gel-electrophoresis: steady state
behavior,
 {\it J. Chem. Phys.} {\bf93}, 9049 (1990).
 \bibitem{Duke90b}
 T. A. J. Duke,
 Monte-Carlo reptation model of gel-electrophoresis: response to field
pulses,
 {\it J. Chem. Phys.} {\bf93}, 9055 (1990).
 \bibitem{WVD91}
 B. Widom, J. L. Viovy, and A. D. Defontaines,
 Repton model of gel-electrophoresis and diffusion,
 {\it J. Phys. I France} {\bf1}, 1759 (1991).
 \bibitem{BMW94}
 G. T. Barkema, J. F. Marko, and B. Widom,
 Electrophoresis of charged polymers: simulation and scaling in a
lattice model of reptation,
 {\it Phys. Rev. E} {\bf49}, 5303 (1994).
 \bibitem{DSV92}
 T. A. J. Duke, A. N. Semenov, and J. L. Viovy,
 Mobility of a reptating polymer,
 {\it Phys. Rev. Lett.} {\bf69}, 3260 (1992).
 \bibitem{BN97a}
 G. T. Barkema and M. E. J. Newman,
 The repton model of gel electrophoresis,
 {\it Physica A} {\bf244}, 25 (1997).
 \bibitem{BN97b}
 M. E. J. Newman and G. T. Barkema,
 Diffusion constant for the repton model of gel electrophoresis,
 {\it Phys. Rev. E} {\bf56}, 3468 (1997).
 \bibitem{BCM96}
 G. T. Barkema, C. Caron, and J. F. Marko,
 Scaling properties of gel electrophoresis of DNA,
 {\it Biopolymers} {\bf38}, 665 (1996).
 \bibitem{Kolomeisky98}
 A. B. Kolomeisky,
 {\it One-Dimensional Nonequilibrium Stochastic Models, Interface
Models, and Their Applications},
 Ph.D. thesis (Cornell University, Ithaca, New York, 1998).
 \bibitem{Evans81}
 K. E. Evans and S. F. Edwards,
 Computer simulation of the dynamics of highly entangled polymers,
 {\it J. Chem. Soc. Faraday Trans. 2} {\bf77}, 1891 (1981).
 \bibitem{bk98}
 G. T. Barkema and H. M. Krenzlin,
 Long-time dynamics of de Gennes' model for reptation,
 {\it J. Chem. Phys.} {\bf109}, 6486 (1998).
 \bibitem{Heukelum2000}
 A. van Heukelum and H. R. W. Beljaars,
 Electrophoresis simulated with the cage model for reptation,
 {\it J. Chem. Phys.} {\bf113}, 3909 (2000).
 \bibitem{Hill98}
 J. M. D. Hill, B. McColl, D. C. Stefanescu, M. W. Goudreau,
 K. Lang, S. B. Rao, T. Suel, T. Tsantilas, and R. H. Bisseling,
 BSPlib: The BSP programming library,
 {\it Parallel Comput.} {\bf24}, 1947 (1998).
 \bibitem{Valiant90}
 L. G. Valiant,
 {\it A bridging model for parallel computation},
 Commun. ACM {\bf33} (8), 103 (1990).
 \bibitem{Bisseling94}
 R. H. Bisseling and W. F. McColl,
 Scientific computing on bulk synchronous parallel architectures, in
 {\it Proc. IFIP 13th World Computer Congress} (North-Holland,
Amsterdam, 1994), Vol. I, North-Holland, Vol. 1, p. 509.
 \bibitem{CarlonUnp}
 E. Carlon, A. Drzewi\'nski, and J. M. J. van Leeuwen,
 Crossover behavior for long reptating polymers,
 {\it Phys. Rev. E} {\bf64}, 010801(R) (2001).
 \end{references}
 \end{document}